# A Survey on Blockchain in E-Government Services: Status and Challenges

Manal Mansour [1*], May Salama[2], Hala Helmi[3], Mona F.M Mursi[4]

[1]Teachning Assistant, Faculty of Engineering, Shoubra, Benha university, Egypt

[2]Assistant professor, Faculty of Engineering, Shoubra, Benha university, Egypt

[3]Professor of Computer Engineering, Department of Computer Science, Faculty of Computers and Artificial Intelligence, Benha University, Egypt, School of Information Technology and Computer Science (ITCS), Nile University, Giza, Egypt.

[4]Professor Emeritus of Computer Engineering, Faculty of Engineering, Shoubra, Benha university.

*Abstract*—**Blockchain technology is referred to as a very secure decentralized, distributed ledger that records the history of any digital asset. It is being used in numerous governmental and private sector organizations across numerous nations. Surveying the current state of blockchain applications and difficulties in e-government services is the goal of this review. Held to the account are use cases for current facilities that use blockchain. Finally, it examines the research gap in blockchain deployment and makes suggestions for future work for additional research.**

1. INTRODUCTION

Blockchain is a digital collection of transactions that are monitored and recorded in a distributed ledger on a decentralized network, meaning that the network has no central controlling authority. The blockchain is made up of discrete data blocks, each of which contains a record of information and is connected to one another chronologically as illustrated in figure 1. The fact that these links cannot be altered helps to build trust in the network. By safeguarding information exchanges as they happen, this ground-breaking system handles them. Decentralizing data collection, storing, and processing while ensuring data integrity and immutability are the goals of blockchain technology.

A blockchain is a digital ledger that stores information in a data structure called a block, making it a database [1]. Databases also store information in tables, but we can't say that a database is a blockchain, despite the fact that a blockchain is a database. Due to variations in their architecture, designs, behaviors, and security, they cannot be used interchangeably. Peer-to-peer technology, or blockchain, means that no single entity controls the data. The abilities are not centralized in one entity because updates to the information are made through a consensus mechanism [2]. So, the powers are not concentrated in one entity and that's why it supports democracy at work. Unchangeable data entries are also a feature of blockchain technology. Each entry in a blockchain has the property of an immutable record since it is protected with a cryptographic hash. While in the database the architecture is a client-server model and the administrator can modify, alter, or even delete all the records. There are few parallels between both, but many researchers argued that when applications are running untrustworthy, blockchains are appropriate, whereas databases are appropriate when performance is more important than security [3].

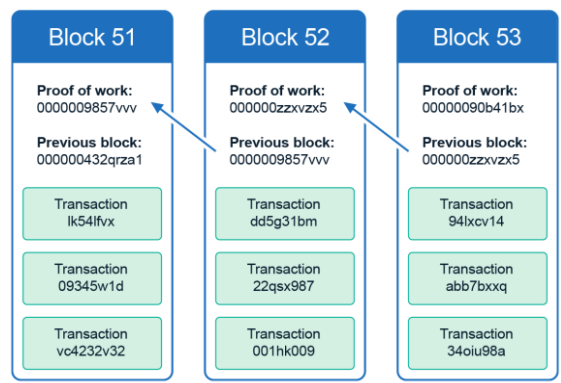

Fig. 1 Blockchain Architecture

The first generation of blockchain technology was initially introduced in cryptocurrency applications [4]. By that time, the world believed that this technology could do more than currency applications. In this sense, Ethereum is a representation of blockchain technology's second generation. Smart contracts [5], are blockchain-based software that executes when certain criteria are met. They are triggered by an event such as an expiration date, or the achievement of a specific price [6].

E-government has investigated blockchain technology, like many other industries, to support the transformation of public administration and to make it easier to deliver transparent and secure public services. Personal data is a common component of e-government services, so it must be carefully protected. In order to respect users' privacy when publishing transactions in the ledger , and take into account legal restrictions like those imposed by the General Data Protection Regulation (GDPR), proposed solutions must also be able to grant public administration parties and other stakeholders the required authorized access. [7].

Blockchain Technology has various limitations, implementation challenges, security challenges, and economic, regulatory, and political challenges [4] [8].

Most of our reviewed papers were published after the year 2016. Figure 2 shows the type of the selected references. The type of references was used to indicate whether the authors are talking according to the research view, the industry view, or both. Most of the selected papers, about 50 (71%), are written by authors that work in different academic institutions, while a small percentage of 7% (5 papers) was from the industry. The remaining 21% (15 publications) were the result of industry and academic partnerships, which argues that because blockchain technology is still maturing, commercial companies may be hesitant to begin integrating it into the marketplace.

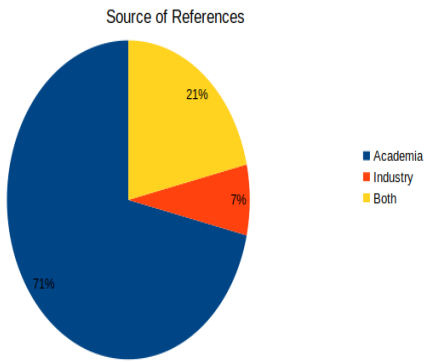

Fig. 2 Source of References

We divide the selected papers into four categories. Survey papers, technical papers for blockchain technology, reported challenges with some suggested solutions for them, and the technical papers for some blockchain-based applications as shown in figure 3. It can be seen that These papers provide information on how to build blockchain applications, however, more work must be done to bring blockchain technology to maturity by creating more applications as the literature's example implementations are insufficient.

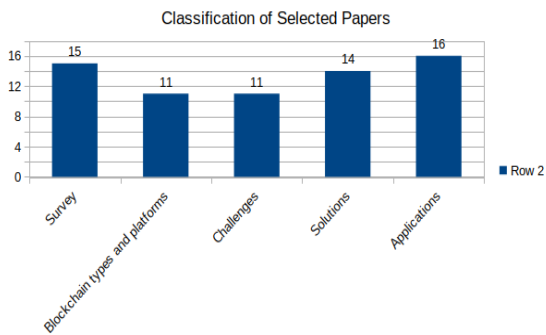

Fig. 3 Classification of References

Since each paper seems to address one or more aspects of the identified blockchain use cases, we selected some identified use cases to further classify the papers as shown in figure 4.

This paper is organized as follows: Section 2 presents an overview of blockchain types and discusses some popular blockchain platforms. Section 3 reviews the challenges and solutions for employing blockchain technology in e-government, while section 4 reviews the current international use cases. A detailed analysis of the collected literature is presented in section 5 which discusses the relevant open issues.

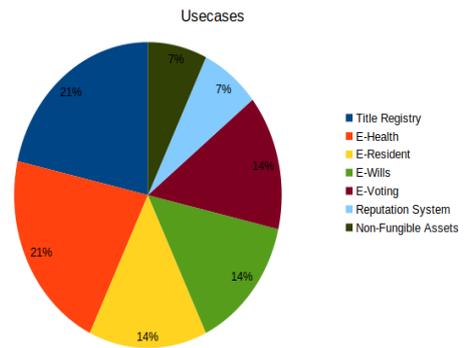

Fig. 4 Distribution of selected use cases

2. TYPES AND PLATFORMS OF BLOCKCHAIN

**2.1 Blockchain Types**

Blockchain comes in three forms [9][10][11]: public blockchain, private blockchain, and consortium blockchain. In a public blockchain, anyone can join and participate. All nodes are equal in rights and have full disclosure of the blockchain system. Private blockchains have strong governance over who has access to what data within the network. In the network, only master nodes are permitted to take part in the validation and verification of transactions. The consortium blockchain, which combines some properties of public and private blockchains, can be viewed as having a partially decentralized structure.

Figure 5[10] shows the different types of blockchains.

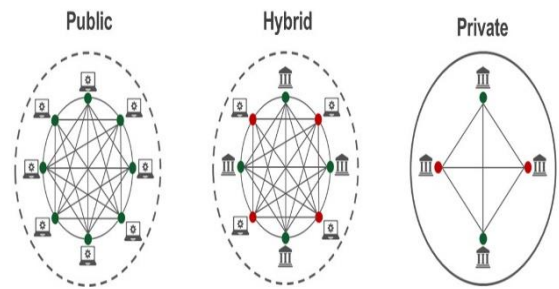

Fig. 5 Public, Consortium, and Private Blockchain

The consortium blockchain is a hybrid between the low trust of public blockchains and the single highly trustable entity model of private blockchains. In table 1 we sum up the major differences. Many studies argued that the consortium blockchain option is the optimum choice for the technical basis of the government information-sharing system. As the public blockchain with full disclosure could reduce the security of the blockchain system to some extent and the closure of the private blockchain results in decreased creativity.

|  | **Public** | **Private** | **Consortium** |
|---|---|---|---|
| Permissionless? | Yes | No | No |
| Who can read? | Anyone | Selected | Selected |
| Who can modify? | Anyone | Approved Participant | Approved Participant |
| Ownership | Nobody | Single entity | Multiple entities |
| Participants known? | No | Yes | Yes |
| Transaction speed | Slow | Fast | Fast |

| Immutability | Impossible | Could be tampered | Could be tampered |
| --- | --- | --- | --- |
| Centralization | Decentralized | Decentralized | Partially decentralized |
| Performance | Low | high | high |

Table 1 Blockchain types

## 2.2 Popular Platforms

Various foundations have collaborated to offer various blockchain platforms; most of which are open source. Deciding which platform to use depends on the requirements. The following is a brief about the most popular blockchain platforms.

BITCOIN: The first well-known and frequently used cryptocurrency in the world, Bitcoin, runs on a peer-to-peer network without the aid of banks or central authority, and is referred to as a first-generation public blockchain. The blockchain network is responsible for managing transactions and issuing currency collectively. Bitcoin uses the PoW [12] consensus protocol, which is energy-intensive, to verify transactions. It was built primarily for financial systems that use public blockchain. it's permissionless and not open for development.

ETHEREUM: Ethereum is an open-source platform where the developer can create decentralized online services on the DAPPs that operate on the basis of smart contracts [13]. Ethereum network can be a public or private network and it uses a Proof of stack (PoS) to reduce power consumption and increase performance. Ethereum also has a built-in Ether coin [6] [14] which is the cryptocurrency generated by Ethereum miners as a reward for computations performed to secure the blockchain.

HYPERLEDGER: It works by offering the essential framework and guidelines for creating different blockchain-based systems and applications that can be used by various enterprise industries. Early in 2016 the Linux Foundation launched the Hyperledger project as an open-source permissioned blockchain [15]. One of its main goals is to create enterprise distributed ledger frameworks and codebases. Over 180 companies from the financial, supply chain, manufacturing, Internet of Things, and technology sectors collaborate with Hyperledger. There are several related initiatives, such as Hyperledger Fabric, Sawtooth, Composer, and Cello. The Hyperledger Fabric Platform is the most well-known and one of the Hyperledger projects that were initially contributed by IBM [16]. Given its modular architecture, which enables plug-and-play components around consensus and membership services, it is an appealing blockchain platform for enterprise solutions.

RIPPLE: It is an enterprise blockchain solution for various payments [17]. It was built in 2012. Through RippleNet, it seeks to link corporations, banks, payment providers, and exchanges of digital assets in order to facilitate nearly-free international transactions free from chargebacks. Through its digital asset "Ripples or XRP," which has grown to be one of the most well-known cryptocurrencies, it facilitates international payments and serves as the medium for banks and other financial institutions to integrate the service into their systems and make it available to their customers. As compared to most other blockchains, XRP is said to be faster and more scalable. One of the quickest digital currencies available today, XRP enables simple global real-time transfers.

QUORUM: Ethereum version targeted at businesses. Quorum is made to handle use cases that call for a permissioned group of players, high-speed, and high-throughput processing of private transactions. Depending on how smart contracts and networks are set up, it can execute hundreds of transactions per second by using vote-based and other algorithms instead of using the Proof of Work (PoW) consensus technique. Alongside Ethereum, Quorum is intended to develop and evolve. Quorum is able to rapidly and easily incorporate the majority of Ethereum changes because it only slightly affects Ethereum's core. In addition, Quorum is open-sourced, completely free to use, and supportive of experimentation.

R3 CORDA: this distributed ledger considered as A revolutionary operating system for the financial services industry used to store and process financial agreements. Established in 2015 by a group comprising some of the largest financial institutions in the world. Corda is considered a permissioned blockchain that limits access to data within an agreement to only those expressly permitted to it. It also lacks a built-in token or cryptocurrency.

Table 2 introduces a comparison between six different platforms and their usability in public services.

### 3. CHALLENGES OF USING BLOCKCHAIN IN PUBLIC SERVICES

#### 3.1 Security

There are many security problems in the blockchain. Some of these security issues are common in the first and second generations of blockchain, and others are specific to the second generation i.e., smart contracts.

**3.1.1 Security Issues Related to Blockchain 1.0**
**The Majority Attack or 51% Attacks** [18]: this attack can be done if many organizations unite to enhance their ability to mine using a proof of work algorithm (POW). If their computing power reaches 51% of the total chain, it can find Nonce value quicker than others. this means this organization has the authority to decide which block is permissible, to reverse transactions or initiate a double spending attack (the same coins are spent multiple times) [19][20].

**Hard Fork:** The problem arises when the system updates to a new version or new agreement and the older versions are incompatible with the new one. Some older nodes decide to continue operating under the previous regulations and chose not to upgrade, so they will operate under a different chain entirely. As with BTC and BTC-cash See figure 6.

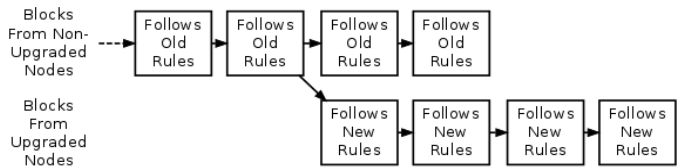

Fig. 6 Hard Fork

| | **Bitcoin** | **Ethereum** | **Hyperledger** | **Ripple** | **R3corda** | **Quorum** |
|---|---|---|---|---|---|---|
| Purpose | Cryptocurrency | Run smart contracts | Industry, use cases | Financial Services | Financial Services | Industry use cases |
| Data stored inside | Cryptocurrency transactions | cryptocurrency, digital assets, smart contracts | Chaincode, smart contracts | Cryptocurrency | Smart contract | smart contract |
| Mode of Peer Participation | Public Networks | Public/Private | Permissioned | Permissioned | Permissioned | Permissioned |
| Consensus Mechanism | Proof of Work | Proof of stack | Pluggable Algorithm: No mining required | Probabilistic Voting | Pluggable Algorithm | Majority Voting |
| Cryptocurrency | Built-in BTC | Built-in Ether | No built-in cryptocurrency | Ripple (XRP) | None | None |
| Throughput | 5 tps | 10 tps | 2000 tps | 1500 Tps | 170 tps | 100 tps |
| Minning Fees | Transaction fees | Gas Fees | No Minning | XRP | No Minning | Ether |
| Language | Script | Solidity | Java, Go | Google's Golang language, C++ | Kotlin | Solidity |
| How can participate? | source code GitHub | source code GitHub | Register for identity to get membership services. | Open source | Open source | GitHub |
| Smart-Contract Functionality | NO | Yes | Yes | NO | Yes | Yes |

Table 2: Comparison between Blockchain Platforms

**Soft Fork:** means that if the system introduces a new agreement or version, and it is backward compatible with the old agreement or version, no nodes need to upgrade to preserve consensus. The issue here is that participants who haven't been upgraded will still believe that incoming new transactions are legitimate. The problem is that the network will reject non-upgraded miners' blocks and efforts when they attempt to mine new blocks. Because members are encouraged to update to the latest version, it can be thought of as a gradual updating method figure 7[20].

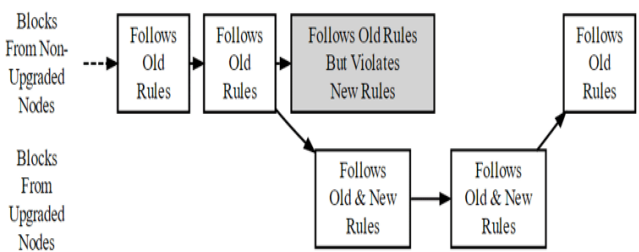

**3.1.2 Security Issues Related to Blockchain 2.0 Transaction-Ordering Dependence (TOD)** [21]: This is the most well-known problem that arises when a single contract invokes several transactions, and the order in which they are executed might have an impact on the chain's new state. The results of miners' efforts alone decide the sequence in which transactions are carried out.

**Re-entrancy vulnerability**: occurs when the smart contract is invoked, changing the actual state of the contract account after the call is finished. The intermediate state can be used by an attacker to repeatedly invoke the smart contract. If the contract is invoked for an e-coin transaction, this could lead to illegal currency theft, double spending, unexpected behaviours, or even the eventual use of all the gas.

**Mishandled Exceptions:** This attack goes after the contract that makes the call; if the called contract encounters an exception, it terminates and returns false, but it may not alert the caller contract. When contract X calls contract Y, in some cases contract X must explicitly check the return value to ensure that the call was correctly executed. If contract Y runs abnormally, it will stop and return false.

**Timestamp dependence:** Every block in the blockchain has a timestamp. The timestamp, which is established by the miner in accordance with its local system time, determines the trigger conditions for some smart contracts. Contracts that depend on timestamps are open to modification by an attacker.

**A denial-of-service attack:** As Ethereum determines the gas value depending on the execution time, bandwidth, memory occupancy, and other factors, an under-priced operation might take place. The gas value often varies in direct proportion to the amount of processing power used by the operation. However, it is challenging to assess the use of computer resources by a specific process, therefore some gas values are not set correctly. For instance, because the gas values for some IO-heavy activities are set too low, these actions can be run in large quantities in a single transaction, allowing an attacker to launch a DoS assault on Ethereum.

**Under-optimized smart contract:** In the second-generation blockchain, the user must pay a fee to have his transaction mined. For instance, an Ethereum network user must pay a fee in the form of ethers to carry out transactions. The code should therefore be optimized. But at least one of these three problems can be found in many contracts. Dead code is code that is present in a contract but will never be executed, despite still using more gas. 2. Expensive loop operations. It refers to some costly procedures that can be performed outside of a loop in order to save gas usage. 3- Opaque predicate: This refers to the running of useless codes that have no impact on operations but nevertheless burn a lot of gas.

### 3.1.3 Security Enhancements

Suggested solutions to deal with the above problems:

**Smart Pool**: In [22] they proposed a novel mining pool system whose workflow is shown in figure 8[22]. The transactions that contain information on mining tasks are obtained by Smart Pool from the client node (i.e., geth). After performing the hashing calculations based on the tasks, the miner sends the finished shares to the smartpool client. They will be committed to the Ethereum smartpool contract when the number of completed shares hits a specific threshold. The shares will be verified by the smartpool contract, and the client will receive rewards.

**Oyente:** It is one of the suggested tools that search for potential security bugs [23][24] Both the bytecode for the smart contract and the current state of Ethereum are the inputs to oyente.

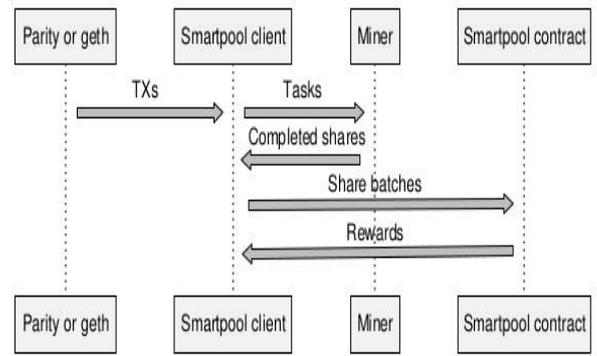

Fig. 8 Smart Pool's execution process

The CFGBUILDER inside oyente will build a control flow graph of the byte code based on the bytes. EXPLORER then simulates the execution of a smart contract in accordance with the Ethereum state and CFG data. CFG will become much more enhanced and improved during this process. The first four separate vulnerabilities listed in part 3.1.2 are discovered by the CORE ANALYSIS module using the associated analysis procedures. The detected vulnerabilities and vulnerable routes are verified using the VALIDATOR module. Finally, confirmed vulnerability and CFG data will be exported to the VISUALIZER module, which users can use for program analysis and debugging. Currently, Oyente is open source for public use as shown in figure 9[23].

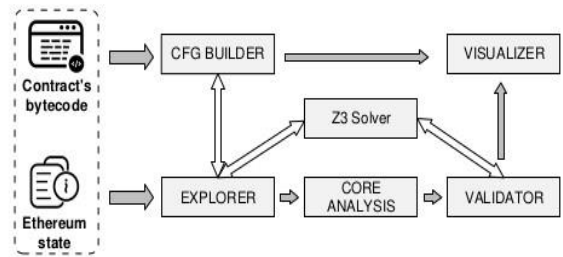

Fig. 9 Oyente's architecture design and execution process

**Town Crier (TC)** is an authenticated data feed system for the data interaction process as Smart contract often needs to interact with off-chain (i.e., external) data source. Since blockchain-based smart contracts cannot contact the network directly, they are unable to obtain data via HTTPS. As seen in figure 10, TC precisely serves as a bridge between smart contracts and HTTPS-enabled data sources. The primary duties of the TC server are to collect data from target HTTPS-enabled websites and to collect data requests from user contracts. In the end, the TC server will send digitally signed blockchain messages back to the user contracts as a datagram. The security of the data request procedure is to a considerable extent protected by TC. For the off-chain data interaction of smart contracts, the TC system offers a strong security architecture, and it has already been made available online as a public service.

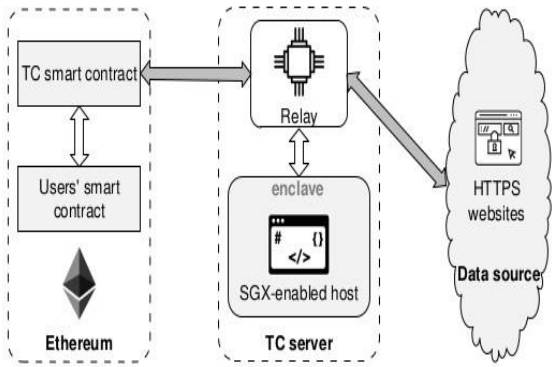

Fig. 10 Architecture of Town Crier system

### 3.2 Privacy

Because a blockchain is distributed, every complete node that processes transactions and creates the blockchain must have access to the actual blockchain transaction data. This means that in a cryptocurrency like bitcoin, the blockchain is accessible to everyone and every transaction can be traced back to the initial genesis block, allowing the connection of a person's several appearances and revealing their identity. Additionally, applications that rely on smart contracts, like e-government systems, may collect, store, and handle a significant amount of confidential data about citizen, clients, employees, products, and research. Users' trust and confidence are typically lost when such information is compromised. Cross-site scripting (XSS) and SQL injection are problems with the present online e-government services due to the lack of proper authentication mechanism applied to input data from user and denial of service attacks (DOA) according to the reports in [25] [26]. Some countries replaced the old systems by blockchain technology; however, the complete view of privacy-preserving is not completely clear. Privacy can be enhanced with respect to identity perspective and data perspective.

#### 3.2.1 Identity Protection

Mixer [27] scrambles multiple flows of transactions. The sender first sends the currency to the mixing provider, who mixes it with other transfers received within a predetermined period. The new scrambled record of transactions is then forwarded to the destination addresses. As a result, the link between the original transactor and their counterpart is broken.

**Zerocoin:** uses Zero-knowledge proof to allow a fully anonymous currency transaction. The coins are minted first and later redeemed with a totally new one that has no history. The participant exchanges their base coin for a zerocoin that has no transaction history at all. It functions as a kind of coin laundry. Minting is the term for this altering process. The participant combines a secret random number r with a coin serial S that will be made public. Combine them and introduce the H(S,r) as a brand-new representation of their currency in the network. By demonstrating that he is aware of r and that h(r,s) corresponds to one of the mints in the blockchain, the participant uses non-interactive ZKP to authenticate coins to prove that they are his coins. The rule of miners here is to ensure that this serial number S hasn't been spent before as in figure 11 [28].

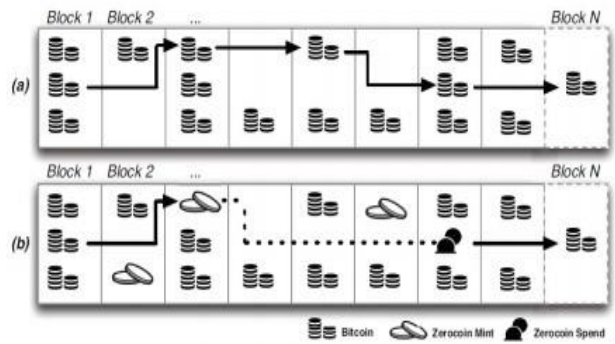

Fig. 11 A typical bitcoin transaction history is shown in chain a, where each transaction is connected to a previous transaction. Chain b shows a chain of zero coins. The linkage between mint and spend cannot be inferred from the block chain data.

**Ring Signature:** [29][30] allows a user (Member of a set) to sign a message on behalf of the "ring" of members but no way to tell who is the one that signed. In order to form a signature on behalf of the group, the signer must input the public keys of all the ring participants (including his own) to the algorithm input and use his own private key as a secret. $PK_0, PK1....PKn$

The message has been signed by one of the sets, but the particular signer is unknown to the verifier in this case. The verifier can identify which member of the ring calculated the signature with a probability of 1 / n. Figure 12[31] illustrate the process of signature calculation.

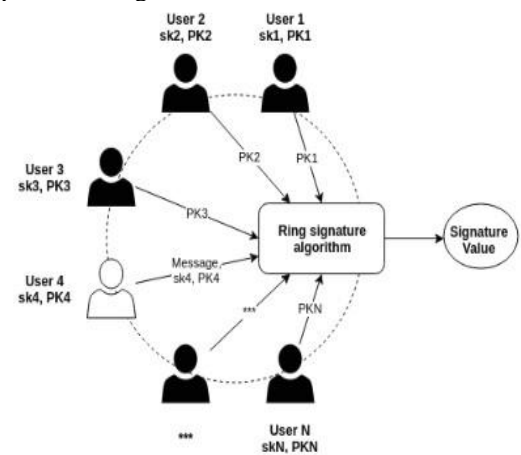

Fig. 12 Signature calculation process

#### 3.2.2 Data Protection

In [32] from the data perspective they proposed a blockchain model called Hawk that does not store the transactions clearly on the blockchain. Developers can separate smart contracts into the private and public portions using the Hawk concept. Codes relating to private information and financial operations may be written in the private portion, whereas codes unrelated to private information may be written in the public portion. Hawk compiles into three parts. (1). the code that, like smart contracts in Ethereum, will run on each virtual machine on each node. (2). the software that can only be run by smart contract users. (3). The manager, a specific trustworthy person in Hawk, will run the program. The manager has access to the contract's private information but will keep it confidential. If the manager terminates Hawk's protocol, it will be automatically financially penalized, and the users will gain compensation.

### 3.3 Scalability

Using blockchain technology in the government services will lead to having a large amount of data stored in the chain whereas the services have to be introduced instantly. There are two problems regarding scalability, throughput, and capacity. The first is generated from the waiting time for any transactions to be mined and included in a block because the block size is limited. Besides, if the block generation time is short, many forks will be generated, so the block generation time cannot be artificially reduced. On the other hand, if all the data are stored in a chain, the size of the chain becomes too large. Current blockchain size of Bitcoin and Ethereum are 163.34GB and 667.10GB consequently. Till 2020 there are several ways to solve these issues [33].

#### 3.3.1 Scalability Enhancements

**Onchain**: The key to the solution is to change only the elements within a block. The bigblock, which essentially increases block size, such as in Bitcoin Unlimited, is one of the fundamental examples of an onchain solution. When compared to the conventional method, this method has the advantages of a high transmission rate and a cheap transmission cost. However, as the propagation speed decreases, forks become more frequent, increasing the likelihood of orphan blocks appearing and the chain's maintenance expenses. In [34] Segregated Witness is proposed. Pieter Wuille created the (SegWit) protocol improvement in 2015. The concept was developed to address the issue of scalability by altering the architecture of the transactions by storing the signatures in a separate data structure that was not computed as a part of the transaction hash or transaction ID. More transactions could be kept in a single block, enhancing the network's transaction throughput, by isolating the witnesses' signatures from the data.

**Off-chain** this carried out by processing the transactions outside of the blockchain, this solution is to increase scalability. Because it keeps the main chain's state intact while also applying the most recent state that was handled by the other channel, this is also known as a state-channel solution. An off-chain solution is something like the Lightning network for Bitcoin. The initial step in this strategy is to establish a channel, like a payment channel, between the two partners. Once the channel is open, the partners can use it to complete a large number of off-chain transactions with no additional mining costs and nearly no waiting time. Opening the channel is an on-chain procedure that costs mining fees. When the channel is closed, the main chain is informed of the final status of the channel.

**Sidechain** enables assets in blockchains like Bitcoin, to be transferred between different blockchains while preventing the assets from malicious attackers and also ensuring the atomicity of the transfers. The protocol which is used to send the assets and call them back to parent-child is called Symmetric Two-Way Peg. This is because, in order to prevent duplicate spending and denial of service, the assets' locking and unlocking actions are followed by a brief time during which the newly moved assets cannot be used [34].

Shard protocol [35] divides the blockchain network into several smaller networks, each contains a part of nodes called a share. Each node only needs to handle a portion of the incoming transactions because the network will split up the processing of transactions into multiple shards. Because different shards can process transactions concurrently, the network's throughput will rise as well as concurrency of transaction processing and verification [36]. it is a critical issue to protect the decentralization and security of the whole system. There are other factors that must be considered, in particular (a) how to obtain consensus on each shard and guard against common threats like double spending and 51% vulnerability. (b) How to handle cross-shard transactions quickly while ensuring the consistency of these transactions.

|  | Issues | Enhancements | Reference |
|---|---|---|---|
| Security 1st Generation | 51% Attack | Smart Pool | [22] |
|  | Soft Forks |  | [19] |
|  | Hard Forks |  |  |
|  | Double Spending |  |  |
| Security 2nd Generation | Transaction o*rdering* | Oyente | [23] |
|  | Dependence (TOD) |  | [37] |
|  | Re-entrancy Vulnerabilities |  | [38] |
|  | Mishandled Exception |  |  |
|  | Timestamp Dependence |  |  |
|  | Denial of services |  |  |
|  | Under-optimized smart contract | Town Crier | [39] |
| Privacy | Identity Revealed | Mixer Zero Coin | [28] |
|  |  | Ring Signature | [30] |
|  | Data Compromise | Hawk | [32] |
| Scalability | Throughput & Capacity | On-chain Solution<br>Off-chain Solution<br>Side-chain solution<br>Shared Protocol | [33] |

Table 3 Blockchain Challenges

## 4. INTERNATIONAL USE CASES

While governments around the world have not fully adopted blockchain technology in the public sector to explore the potential of blockchain technology in providing people with public services, numerous nations have started to try some solutions using blockchain projects. Until now, a reliable third party is required for the settlement of the market. Hence, the great potential of this technology is evident. Direct transactions are made possible by Blockchain, which also eliminates the need for a central actor that might have previously controlled the data, received a commission, or even interjected to censor it. Blockchain applications can be divided into two main categories: the financial sector which is related to concurrency projects, and the public sector Which attempts to employ this technology to provide social services electronically [40].

### 4.1 Blockchain in Financial Sector

One industry with clear applications for the blockchain is financial services [41]. Blockchain holds the promise of bringing greater efficiency and transparency to many financial applications by adopting both public and private blockchain types. BTC is one of the cryptocurrencies which uses a public blockchain platform. Since banks and governments do not issue or support bitcoin, only balances are recorded on a public ledger that is transparently accessible to all users. Despite most cryptocurrencies are not legal tender in most parts of the world, there are several cryptocurrencies now, BTC, ETH, XRP., etc. The most well-known usages of cryptocurrency are sending and receiving payments at low cost and high speed and investing in innovative early-stage start-ups. On the other hand, the banking industry introduces an example of using private blockchain technology, by allowing cross-border transactions to be made in real-time and money to be exchanged at the speed with which information moves today. Introducing this technology to applications like banking will improve transparency, add security, and lower cost. Corda is a blockchain platform that is directed basically toward banking services. HSBC bank is the first bank to apply corda enterprise blockchain published in Google Cloud [42]. The potential problems with blockchain in the financial sector are that the traditional financial institutions make money on transaction fees that could be lowered or eliminated with blockchain technology. Therefore, banks could face challenges in volume and transaction-based revenue. Another obstacle in incorporating blockchain in financial services is that regulation hasn't caught up yet, because of the speed of progress of blockchain innovation.

### 4.2 Blockchain in Public Services

#### 4.2.1 Title Registry

Brazil, the Republic of Honduras, the Republic of Georgia, Ukraine, the United Arab Emirates, and Brazil were among the first countries to implement Blockchain technology to enhance their property register. However, there is currently no one preferred framework; instead, hybrid Blockchain solutions are frequently used as a step solutions. National Agency of Public Registry (NAPR) published the project for Georgia's use case. Their solution stored the crucial records on a private blockchain and then, using the public Bitcoin blockchain to publish hashes of essential documents. This hash is a fixed-length bit string generated from variable length input which is the unique register or document, then this hash is posted on a public blockchain.

This strategy is believed to guarantee user data privacy and make use of the immutability of public blockchains. The timestamped documents are made available in the public chain[43]

. The full documents and associated transactions being stored are placed in a NAPR backend database on the private chain. So, citizens still need to visit NAPR offices to complete transactions.

VeriDoc and provdoc were projects published in 2018 and 2019 respectively for identifying ownership of the particular asset, parties involved in transactions and date of signing etc. They are providing a solution for reducing the creation of fake documents VeriDoc solution is designed using multichain [44]. It works as follows: the issuer produces a land title then the system gives the land a unique hash and unique ID. it then embeds that hash as a QR code in a land title. The true land title will be loaded. after reading the QR code, the system will check the hash inside the QR code and connect to the blockchain to search for the true land title to load it.

In [45] some problems faced the title registry project such as double spending where they take it from another perspective; when the person sold the land then sold the building on the land. Every asset shall have a birthdate, asset ID, size/value (depending on asset type), and GPS-based location. First, they divided the types of assets into two parts: divisible and indivisible assets. The divisible assets must be of type leaf, root or sub-root as shown in figure 13[45].

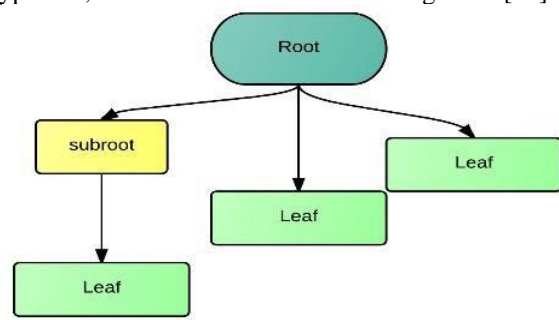

Fig. 13 Divisible Assets

The only available properties for sale are the leaves nodes. If a property, for example, were to be divided into 8 plots, the original land would be transformed into a root asset, and the 8 plots that were created would be given new identities and become leaf assets. The issuing authority will be the government or the producer of the item being traded on the smart contract platform. This is necessary in order to establish the authenticity of the asset which will form part of the smart contract.

According to the authors in [46], there are four key technical requirements that must be met before blockchains can be completely implemented in a land registry: (1) a method for digitizing registers; (2) an identity solution; (3) multiple signature wallets; and, ultimately, (4) raising public awareness.

### 4.2.2 E-Health

Blockchain technology can support the health sector in many applications such as improving medical record management, enhancing the insurance claim process, or accelerating clinical/biomedical research. In [47] they discussed the potentials of using BC technology compared to distributed database and found that BC can introduce (1) decentralized management, (2) immutable audit trail, (3) data provenance, (4) robustness/availability, and (5) security/privacy. The study highlighted some challengat faced this type of application The first challenge is related to transparency and confidentiality. The second was related to speed and scalability. The last challenge is the threat of a 51% attack. According to [48], the aforementioned challenges can be lessened by carefully planning and implementing the biomedical/healthcare application systems. Encrypt sensitive data on the blockchain network, store sensitive data off-blockchain, and only distribute "pointers" (such as encrypted links) or permission information on-blockchain in order to mitigate the transparency confidentiality issue. Smart contracts can also be used to automate data management protocols.

Additionally, viable solutions to the speed and scalability problems include using blockchain as an index of health data rather than a repository for all information and storing only current, validated transactions on the blockchain rather than the entire history. Additionally, compared to the Bitcoin blockchain, many blockchain implementations, such as Big chain DB, offer much faster transaction speeds. A 51% attack is less likely since it uses permissioned blockchain networks, which prevent hostile nodes from joining the network.

### 4.2.3 E-Residency

Estonia opened its digital borders to allow anyone in the world to apply to become an e-resident in December 2014, making it the first country to do so. It does not provide a right to physically enter Estonia or the European Union (EU) however e-residents can remotely access and use a range of Estonian e-government and private sector services e.g., many services, including business and company registration, creating bank accounts, money transfers, buying and selling real estate and other property, and trading products and services are all made possible by the e-ID. In [49], the authors mentioned that they had to deal with the main issue using blockchain which is the use of an inaccurate, inauthentic digital identity. The Estonian e-residency application process deviates from the security standard in three ways: it no longer involves an in-person face-to-face interview with the applicant; it doesn't demand the production of a variety of original documents to substantiate identity; and, finally, it only requires a photo or scan of the identity document used to submit the application. The international standard, Estonian law, and generally recognized banking standards are all violated by taking this procedure for creating a bank account.

### 4.2.4 E-Wills

In [50], they proposed an e-wills system that allows the user to decide how his digital assets should be distributed after his death. The main two phases in their application are 1- Creating the will, start by uploading the will into IPFS (InterPlanetary File System) and a hash to its location saved in the smart contract. A death flag in the contract that will only be set after confirming its existence on a government website. The generated smart contract is subsequently deployed into the blockchain, where it is safe and immune to modification. 2. Probate a will: By entering the testator's public key and a valid death certificate ID, any beneficiary may probate a will. The smart contract will be triggered after the system determines whether a Will exists for the public address entered, resulting in the completion of all transactions in a matter of minutes.

### 4.2.5 E-Voting

In [51] they proposed an electronic voting system for a university campus using private blockchain to prevent the system from any threats within the communication link using the main properties of blockchain which are decentralized property, hashing, and encryption concept for providing security. During the voting process, the vote will be encrypted by the public key of the University Election Commission (UEC). Then signed by the voter's private key. Once the block is completed it can't be altered.

In [52], the authors proposed a decentralized system to be integrated with the voting system to avoid problems like missing names in the voter list and misplaced votes. Their proposal was built using the Ethereum platform and solidity language. This project is suitable for small-scale elections in associations and small towns and does not fit perfectly the general political elections [53].

### 4.2.6 Reputation System

Reputation is a technique to measure how much the community would trust an individual and can be calculated by considering one's previous transactions and interactions with the community. The greater one's reputation the more trustworthy he is regarded by others. In [54] a blockchain-based distributed system for educational records and reputation was proposed. First, an initial award of educational reputation currency would be issued to each institution and intellectual worker. A staff member may receive an award from an institution if some reputation records are transferred to them. All reputation changes might be easily tracked because transactions are kept on the blockchain. They argue that it might contribute to opening up the system of scholarly reputation that is now attached to academics. In [55] proposed a reputation model based on blockchain for web community, in which a voucher will be signed if customer is satisfied with the service. To prevent a sybil attack, a service provider must deduct an additional 3% of the payment made to the network after signing a voucher.

#### 4.2.7 Non-Fungible Assets (NFA)

NFT [56] originates from an Ethereum token standard that aims to distinguish each token with recognizable signals making it suitable for identifying something or someone in a unique way. It cannot be exchanged like bitcoin or any other coins. By utilizing NFTs on smart contracts, a creator can easily prove the existence and ownership of digital assets in the form of videos, images, works, a piece of art, event tickets, or any other type of digital asset. Additionally, the creator also can earn royalties for each profitable transaction made on any NFT market. Although NFT is in a very early stage due to its complexity, it offers great opportunities in many fields. (1) Game industry is one of these fields in which both players and developers profit from the NFT markets. Now there are already some crypto games CrytpoKitties, Cryptocats, CryptoPunks with fascinating features to attract a lot of investors to join the game e.g., Making unique characters or getting the limited-edition virtual pets, then reselling them at a high price. (2) Flourishing virtual events and protecting digital collectibles, where the users can buy and sell the digital assets using smart contracts without relying on a third party.

An overview of blockchain use cases and applications may be seen in table 4. Because it is impossible to gather information on all of the current use cases, this table should not be considered complete. New blockchain solutions occur on a regular basis as a result of the rapid growth of the technology. The table below is designed to provide an overview and broad concept of the most often utilised solutions today. Some of them are still in the testing stage, so they don't have complete functionality [57].

#### 5. DISCUSSION AND OPEN ISSUES

Identifying the research gaps will help other researchers to focus their work on the areas that are still open and will help to understand unanswered questions in current Blockchain technology.

1- The authors in [58] observed that topics like latency, throughput, size and bandwidth, versioning are rarely taken in the current literates. No researchers are studying the usability of this technology from the developer perspective i.e. the difficulty of using Bitcoin API has not been tackled yet. This could spark more applications and solutions to the blockchain environment.

2- As this technology is considered as an emerging field there are a lot of models, frameworks, and proposals that need to be proven by testing, so there is a crucial need for researchers to develop more prototypes and proof-of-concept to deepen the understanding and maturity of this technology to improve stakeholders' confidence in the use of this technology and to foster its adoption in nonfinancial applications. All the discussed benefits and obstacles should be proven for each case and addressed through impact analyses and thorough examination of real applications.

3- There is a need for open standards to achieve the interoperability of multiple blockchains.

While blockchain technology will reduce the need to trust anyone of the individual actors, it does not eliminate the need for trust because blockchain-based systems are considered socio-technological assemblages that are made up of a wide variety of actors [59], including miners, validators, programmers, token holders, and end users. In order to restore confidence among users, good governance policies must be developed to stop these actors from acting in an untrustworthy manner. [60].

#### 6. CONCLUSION

Blockchain technology has proved its potential for transforming many aspects of the traditional services with its key features: decentralization, audibility, persistence, and anonymity. In this paper, a comprehensive survey on the use of blockchain in e-government services is presented. Challenges and problems that would hinder blockchain deployment in such services were investigated along with some approaches for solving these obstacles. Our findings show that governance model, Storage criteria and Access control are the main problems we have to tackle when using blockchain as the infrastructure for building government model. Future studies should concentrate more on BC solutions to address the aforementioned problems and use a critical evaluation of BC technology adoption's potential advantages in the public sector. Additionally, to explore the different potential benefits, robust research methodologies should be implemented in empirical studies in the context of government. Many novel applications are difficult to implement because test cases still have a lot of flaws and limitations.

| Service | Country | Platform | Problem | Solution |
| --- | --- | --- | --- | --- |
| Title registry | Bitfury for Georgia | Exonum, a private blockchain | Quality of Data | Enhance the property registration mechanism |
| | Estonia | Bitcoin | ownership proven | certificates is timestamped, hashed in bitcoin platform authorized by NAPR. |
| | | | Privacy | Special data structure, Merkel Tree to pool and aggregate hashes. |
| E-Health | United-State | Ethereum | Scalability | Encrypted health data stored off-chain |
| | Estonia | Hyperledger Fabric | protect patients' privacy | Permission blockchain with different rules handles the patients' data. |
| | | BitCoin | performance | Only some nodes are permitted to participate in the |

| | | | | consensus and validation processes |
|---|---|---|---|---|
| E-Residency | Estonia | cyber-security | authentication | E-ID card |
| | Portugal Singapore | | Scalability | X-road Platform |
| E-Wills | Indiana | Ethereum | immutability | N/A |
| E-Voing | Ukrania | Ethereum | Identity Proven | Public-Private key |
| Banking Systems | England-HSBC | Ripple | Privacy,Security | N/A |

Table 4 Summary of some Blockchain Applications

## abbreviations

| | |
|---|---|
| BC | BlockChain |
| BTC | Bitcoin |
| PoW | Proof of Work |
| PoS | Proof of Stack |
| DAPPs | Decentralized Applications |
| NAPR | National Agency of Public Registry |
| TOD | Transaction-Ordering Dependence |
| CFG | Control Flow Graph |
| TC | Town Crier |
| DOA | Denial of service Attacks |
| XXS | Cross-site scripting |
| ZKP | Zero Knowledge Proof |
| SegWit | Segregated Witness |
| NFA | Non-Fungible Assets |
| IPFS | InterPlanetary File System |

**Declarations:**

ACKNOWLEDGEMENTS:

I would like to thank my professors and colleagues for their continuous support.

**Availability of data and materials:**

Details related to data and simulation results are provided in the manuscript. They are not stored in any publicly available repositories.

**Authors' contributions**

This work introduces a comprehensive survey to be a complete guide for any blockchain developer. It will help identify the main challenges and solutions which appear when adopting this technology in public services.

**Competing interests :**

The authors declare that they have no competing interests.

**Funding**:

This work is supported by faculty of engineering shoubra, Benha university,Egypt.